\documentclass[12pt,preprint]{aastex}

\shorttitle{Massive Close Binary HD~101131}
\shortauthors{Gies et al.}

\begin{document}

\received{2002 February 27}
\accepted{2002 April 2}

\title{Tomographic Separation of Composite Spectra. X. \\ 
The Massive Close Binary HD~101131} 

\author{Douglas R. Gies\altaffilmark{1}}
\affil{Center for High Angular Resolution Astronomy and \\
 Department of Physics and Astronomy,\\
 Georgia State University, Atlanta, GA 30303; \\
 gies@chara.gsu.edu}

\author{Laura R. Penny\altaffilmark{2}}
\affil{Department of Physics and Astronomy, \\
 College of Charleston, \\
 Charleston, SC 29424; \\
 pennyl@cofc.edu}

\altaffiltext{1}{Guest Observer,
Mount Stromlo and Siding Springs Observatories, Australia}

\altaffiltext{2}{Guest Observer,
Complejo Astronomico El Leoncito (CASLEO), 
San Juan, Argentina} 

\author{Pavel Mayer}
\affil{Astronomical Institute, Charles University, \\
V Hole\v{s}ovi\v{c}k\'{a}ch 2, CZ-180 00 Praha 8, Czech Republic; \\
mayer@cesnet.cz} 

\author{Horst Drechsel and Reinald Lorenz}
\affil{Dr.\ Remeis-Sternwarte Bamberg, 
Astronomisches Institut der Universit\"{a}t Erlangen-N\"{u}rnberg, 
Sternwartstrasse 7, D-96049 Bamberg, Germany; \\
drechsel@sternwarte.uni-erlangen.de, 
lorenz@sternwarte.uni-erlangen.de}

\slugcomment{2002, ApJ, in press}
\paperid{55623}


\begin{abstract}
We present the first orbital elements for the massive close binary 
HD~101131, one of the brightest objects in the young open cluster
IC~2944.  This system is a double-lined spectroscopic 
binary in an elliptical orbit with a period of $9.64659 \pm 0.00012$ 
days.   It is a young system of unevolved stars (approximately 
2 million years old) that are well within their critical Roche surfaces. 
We use a Doppler tomography algorithm to 
reconstruct the individual component optical spectra, 
and we apply well known criteria to 
arrive at classifications of O6.5~V((f)) and O8.5~V for the 
primary and secondary, respectively.   We compare 
the reconstructed spectra of the components to single-star  
spectrum standards to determine a flux ratio of 
$f_2/f_1 = 0.55\pm 0.08$ in the $V$-band.  
Both components are rotating faster than synchronously.  
We estimate the temperatures and luminosities of the 
components from the observed spectral classifications, 
composite $V$ magnitude, and cluster distance modulus. 
The lower limits on the masses derived from the orbital 
elements and the lack of eclipses are $25 M_\odot$  and 
$14 M_\odot$ for the primary and secondary, respectively. 
These limits are consistent with the larger masses estimated 
from the positions of the stars in the Hertzsprung-Russell 
diagram and evolutionary tracks for single stars. 
\end{abstract}



\section{Introduction}                              

The hot star HD~101131 (HIP~56726, LS~2420) is found 
in the heart of the young open cluster IC~2944.
This cluster contains a population of some dozen 
relatively unevolved O-type stars \citep{wal87,mer95}
\footnote{See also 
http://obswww.unige.ch/webda,  
a Web site devoted to stellar open clusters created by
J.-C. Mermilliod.}
that form the ionizing source of the surrounding H~II region, 
Gum~42 \citep{geo00}.  The two brightest members of the cluster are 
HD~101131 (O6~V ((f)); \citet{wal73}) and its nearby companion, 
HD~101205 (O7~IIIn ((f)); \citet{wal73}).  HD~101131 appears single 
in speckle interferometric observations \citep{mas98}, but 
the spectroscopic evidence to date suggests that it is a close, unresolved binary. 
\citet{fea55} were the first to suspect its spectroscopic binary nature, 
and they observed double-lines on several occasions.
\citet{tha65} reported on new radial velocity measurements 
and plans to determine the orbit, but, unfortunately, subsequent
results were never published.   The system has been relatively 
neglected since, save for a single radial velocity measurement 
by \citet{con77} and one double-lined observation made with the 
{\it International Ultraviolet Explorer Satellite (IUE)} 
\citep{pen96,how97,sti01}. 

We have explored the properties of a number of massive binaries 
in this series of papers using UV spectra available in the 
{\it IUE} archive.   However, in a recent paper \citep{pen02}, 
we applied many of the same techniques to new optical spectra 
obtained from the CASLEO and Mount Stromlo Observatories in 
1997 and 1998, respectively. 
HD~101131 was also a target during these runs, and 
here we present the first orbit for this massive binary.   
We discuss the observations and 
radial velocity measurements for this double-lined system in \S2. 
We then apply a version of the Doppler tomography algorithm 
to reconstruct the individual spectra of both components, 
from which we determine their spectral classifications, 
projected rotational velocities, and flux ratio (\S3). 
Finally, we discuss the probable masses and evolutionary state of 
the binary in \S4.  


\section{Observations and Radial Velocities}        
  
Our spectra were obtained mainly in two observing runs 
at different sites (see \citet{pen02} for details).  
The first set was obtained with the 2.15-m telescope of the 
Complejo Astronomico El Leoncito (CASLEO) and REOSC echelle spectrograph
during the period 1997 March 19 -- 28.  The detector was a TEK $1024\times 1024$
CCD with 24$\mu$m square pixels.  We extracted 23 orders
from these echellograms, and the resulting spectra span the range
from 3575 to 5700 \AA ~with a spectral resolving power of 
$\lambda / \Delta \lambda =$ 13,000 (with a 
signal-to-noise ratio of $\approx 150$ per pixel 
in the better exposed portions of the spectrum).  

Our second observing run took place at the 74-inch telescope 
at Mount Stromlo Observatory over the period 1998 April 6 -- 14. 
These spectra were made with the coude spectrograph and a 
SITe CCD detector (D14) with 15 $\mu$m square pixels in a $4096\times 2048$
format.  The resulting single order spectra cover the range 3804 -- 4220 \AA ~with a
reciprocal dispersion of 0.10 \AA ~per pixel and a resolution
element of 0.30 \AA ~FWHM ($\lambda / \Delta \lambda = 13400$ at 4012 \AA ).   
The MSO spectra have a typical S/N equal to 160 pixel$^{-1}$ in the continuum.

The spectra were reduced
using standard routines in IRAF\footnote{IRAF is distributed by the
National Optical Astronomy Observatories, which is operated by
the Association of Universities for Research in Astronomy, Inc.,
under cooperative agreement with the National Science Foundation.}.
The extracted orders of the CASLEO echellograms 
were rectified to a unit continuum by fitting a high order spline 
function to line-free regions.  
Small amplitude irregularities related to this 
fitting of the echelle blaze function were evident in the
continuum, and the same pattern was seen in all spectra made on a
given night.  We were able to remove most of the pattern
by dividing the target spectrum by a correction spectrum formed from
spectra of B-star, $\tau$~Sco, which was also observed each night.
The correction spectrum was a smoothed version of the particular
night's $\tau$~Sco spectrum divided by a global average
representation of this star's stellar spectrum.
The spectra from each run were collected and  
transformed onto their respective heliocentric wavelength grids.

A visual inspection of the spectra immediately showed evidence
of cyclic variation in profile shape over a period of approximately 
10~days.  However, the binary components were only cleanly separated 
in the \ion{He}{1} $\lambda\lambda 4026,4471$ profiles 
of spectra obtained near the orbital quadrature phases.  
Thus, we decided to measure these generally blended features 
using a scheme of fitting single-star template spectral lines 
to our observed composite spectra \citep{pen02}. 

We used the star HD~57682 (O9~IV; \citet{wal72}), 
which was observed in both runs, as the template for the fainter 
secondary, but, unfortunately, we observed no hot star in 
common between the CASLEO and MSO runs that could be used as 
a template for the primary.   We selected 
HD~93160 (O6~III(f); \citet{wal72}) as the primary star template 
for the CASLEO spectra, and we used HD~164492 (O7~III((f)); \citet{wal72})
as the template for the MSO spectra.  All of the template spectra 
stars appeared to be constant-velocity objects in our observations. 
The individual spectra for each template were averaged together 
to improve the signal-to-noise ratio (S/N), shifted to the rest frame, 
and any obvious interstellar lines were excised.   
The radial velocities of the template stars were measured by parabolic  
fitting of lines taken from the list of \citet{bol78}, 
and we found average radial velocities of $+25.0 \pm 0.5$ and 
$+26.0 \pm 1.1$ km~s$^{-1}$ from the CASLEO and MSO spectra, respectively, 
for HD~57682; $-1.0\pm 2.3$ km~s$^{-1}$ for HD~93160 (CASLEO); and 
$+5.3\pm 0.9$ km~s$^{-1}$ for HD~164492 (MSO).  

Next, we artificially broadened each template spectrum 
by convolution with a rotational broadening function 
to produce profiles that matched the spectral components
of HD~101131 in the best-separated quadrature spectra.  
We also used these resolved profiles to estimate the 
line-depth ratio between the components.   Once these
fitting parameters were set, we determined the radial velocities
of each component for a given line by a least-squares fit of
the observed profile with the co-addition of the two template profiles
shifted in wavelength to obtain the best match.
We used this technique to measure radial velocities
for the strongest lines in the CASLEO spectra, specifically
\ion{H}{1} $\lambda\lambda 3797$, 3835, 3889, 3970, 4101, 4340, 4861; 
\ion{He}{1} $\lambda\lambda 4026$, 4471, 4713, 5015; 
\ion{He}{2} $\lambda\lambda 4199$, 4541, 5411;
and \ion{O}{3} $\lambda 5592$.  A subset of these
(plus the \ion{C}{3} $\lambda 4070$ blend) was measured in 
the smaller wavelength range of the MSO spectra. 

There was no evidence of systematic line-to-line 
differences in the radial velocity measurements, and so 
no line-specific corrections were applied. 
The radial velocities from all the available lines were averaged 
together after deletion of any very discrepant measurements. 
Finally, we made small adjustments to these averages based 
on measurements of the strong interstellar \ion{Ca}{2} 
$\lambda\lambda 3933, 3968$ lines.   
An interstellar spectrum was formed by extracting the
mean spectrum in the immediate vicinity of each interstellar
absorption line.  
(We made Gaussian fits of the interstellar \ion{Ca}{2} profiles 
in the average spectra, and we found the radial velocity 
was $-6.4\pm 0.2$ and $-4.9 \pm 0.3$ km~s$^{-1}$ for the
CASLEO and MSO spectra, respectively.)  
We then cross correlated this spectrum with
each individual spectrum to measure any small deviations in
our wavelength calibration (generally $<2$~km~s$^{-1}$),
and these small corrections were applied to the mean velocities. 
Table~1 lists the heliocentric dates of mid-observation, orbital phase, 
and, for each component, the mean radial velocity, the standard 
deviation of the mean, the observed minus calculated ($O-C$) residual 
from the orbital fit, and the number of lines used in the mean. 

We also obtained 8 high resolution CCD spectra between 1992 and 1994 
using the European Southern Observatory (ESO) 1.52-m telescope, 
equipped with the ECHELEC echelle spectrograph, and the ESO 1.4-m
coud\'{e} auxiliary telescope feeding the coud\'{e} echelle spectrograph
of the ESO 3.6-m telescope \citep{lor99}.  Radial velocities measured 
from these spectra are listed in Table~1 along with data from \citet{fea55},  
\citet{con77}, and \citet{sti01}.  
After publishing the radial velocities of HD~101131 in 1955, the Radcliffe
Observatory staff obtained more spectra.  These were made available to one
of us (PM); due to the low dispersion of these plates and the wide spectral lines, 
only the primary component (or the blended feature) was visible. 
We measured radial velocities by matching the line profile with 
its wavelength reversed version \citep{lor99}.  
Usually only H$\gamma$ and H$\delta$ were measurable, but in some cases 
results for several \ion{He}{1} lines were included in forming the 
mean radial velocity (listed in Table~1).   

\placetable{tab1}      

We made a preliminary period search using a version of the
discrete Fourier transform and CLEAN deconvolution algorithm
of \citet{rob87} (written in 
IDL\footnote{IDL is a registered trademark of Research Systems, Inc.}
by A.\ W.\ Fullerton), and the position of the strongest peak
in the CLEANed power spectrum for the 
primary velocities was used as the starting value for 
a general solution of the orbital elements using the 
nonlinear, least-squares fitting program of \citet{mor74}.  
We found that the scatter in the pre-1995 measurements 
was significantly larger than in CASLEO and MSO results
and that these early data were often affected by pair blending 
with the secondary's lines, so we made a final fit by assigning 
a lower weight (0.10) to the earlier data.   
The secondary radial velocity measurements have larger associated 
errors, so we fitted these for the systemic velocity, 
$V_{0}$, and semi-amplitude, $K$, only with the other 
elements set by the solution for the primary.   
The solutions are listed in Table~2, and the combined radial
velocity curve is illustrated in Figure~1. 

\placetable{tab2}      

\placefigure{fig1}     

We find residuals from the fit that are approximately $18\%$ of the 
projected rotational velocities (\S3), which is not unexpected 
given the severe line blending in most of the spectra. 
The system apparently has a significant eccentricity, but 
this is not unusual for binaries with similar periods \citep{mas98}.   
The systemic velocities of the two components are approximately 
$3 \sigma$ apart.  The primary's systemic velocity is 
more negative than the secondary's value, which is what we 
would expect due to line formation in the 
greater outflow from photosphere to wind 
in the more luminous primary.   \citet{tha65} find a mean radial 
velocity of $+1.1$~km~s$^{-1}$ for the bright, hot stars in 
IC~2944, a value which sits comfortably between our results 
for the systemic velocities and which is probably close to the 
physical systemic velocity of the binary. 


\section{Tomographic Reconstruction}                

We used the Doppler tomography algorithm described by 
\citet{bag94} to reconstruct the individual primary and 
secondary spectra independently from the CASLEO and MSO 
spectra.  We took the radial velocity shifts for each component 
from the orbital solutions in Table~2, then the reconstruction 
was run for 50 iterations with a gain of 0.8 (the results 
are insensitive to both parameters).  The reconstructed 
spectra are plotted in in Figure 2 in a format similar 
to that used in the spectral atlas of \citet{wal90}.  The 
reconstructions from the MSO spectra are shown just above those from 
the CASLEO spectra (in the short wavelength portion of Fig.~2), 
and there is good agreement between these two sets of spectra.

\placefigure{fig2}     

We compared the reconstructed spectra with the spectrum 
standards in the atlas of \citet{wal90} to determine the 
spectral classifications of the components.   
The strengths of the \ion{He}{1} $\lambda\lambda 4026,4471$ 
lines relative to those of \ion{He}{2} $\lambda\lambda 4200, 4541$
are all consistent with a spectral type of O6.5 for the primary. 
In particular, the weak but convincing presence of \ion{He}{1} $\lambda 4387$
is more consistent with the O6.5 type than the O6~V((f)) type given 
for the composite spectrum by \citet{wal73}.  
The small ratio of \ion{Si}{4} $\lambda 4088$ 
to \ion{He}{1} $\lambda 4026$ 
indicates a main sequence class, as does the near equality of  
\ion{He}{2} $\lambda 4686$ and \ion{He}{2} $\lambda 4541$. 
Thus, we classify the primary as type 
O6.5~V((f)), where the suffix indicates the presence of 
weak emission in \ion{N}{3} $\lambda\lambda 4634-4640-4642$.   
We compare its spectrum in Figure~2 to that of 
HD~93146, which is given as the standard of this class  
in \citet{wal90}. 

The secondary, on the other hand, has features indicating a 
cooler temperature and later type.  All the \ion{He}{1} 
transitions appear stronger in the secondary's spectrum, 
and the defining ratios of 
\ion{He}{2} $\lambda 4541$/\ion{He}{1} $\lambda 4387$ and 
\ion{He}{2} $\lambda 4200$/\ion{He}{1} $\lambda 4144$
are most consistent with a type of O8.5. 
The relative strength of the  
\ion{Si}{4} $\lambda\lambda 4088,4116$ lines compared to the 
neighboring \ion{He}{1} $\lambda\lambda 4121,4144$ features
indicates a luminosity class V, as does the robust strength 
of \ion{He}{2} $\lambda 4686$.  Figure~2 
illustrates the good agreement between the spectrum of the 
secondary and that of HD~46149, which \citet{wal90} use as 
a standard for type O8.5~V. 

The two spectral standards, HD~93146 and HD~46149, provided 
us with the means to estimate the visual flux ratio, 
$r=F_2/F_1$, by matching the line depths in the reconstructed
spectra with those in the standards.   This was done by 
aligning the reconstructed and standard spectra, 
adjusting for differences in the placement of the continuum,
Gaussian smoothing of the spectra to eliminate 
differences in projected rotational velocity and instrumental 
broadening, and then finding a best-fitted line ratio that 
allocates a proportion of flux to each component to 
best match the line depths.   We found $r=0.52 \pm 0.12$ and 
$0.56 \pm 0.12$ for the MSO and CASLEO reconstructions, respectively,  
and we adopted the mean result, $r=0.54 \pm 0.08$, for the 
depictions of the reconstructed spectra in Figure~2. 

Finally, we used the reconstructed spectra 
to estimate the projected rotational velocities of the 
components.   Our procedure involved calculating a 
grid of rotational broadening functions for a linear 
limb darkening law \citep{wad85,gra92} and then  
convolving an observed narrow-lined spectrum with 
these broadening functions.   
We focused on the \ion{Si}{4} $\lambda 4088$
profile for the primary's spectrum since it represents the strongest 
metallic line (intrinsically narrow) in the range covered 
by the MSO spectra.   We compared the spectral reconstructions 
from the CASLEO and MSO spectra with broadened versions of 
the spectrum of the narrow-lined star HD~164492 
($V\sin i = 48$ km~s$^{-1}$, \citet{pen96}).    
The best fitting profile matches with the primary's spectrum
were made with $V\sin i = 102\pm 10$ km~s$^{-1}$. 
The secondary's spectrum shows stronger \ion{He}{1} 
features in the MSO range, so we fit all the profiles in 
the range between \ion{He}{1} $\lambda 4009$ and \ion{Si}{4} $\lambda 4088$. 
Here we compared the secondary's spectrum with broadened 
versions of the spectrum of the star HD~57682 
($V\sin i = 33$ km~s$^{-1}$, \citet{pen96}),  
and the best fits were made with $V\sin i = 164\pm 12$ km~s$^{-1}$.
These are comparable to the estimates from the {\it IUE}
observation \citep{pen96,how97,sti01}. 


\section{Discussion}                                

The mass limits derived from the orbit, $M \sin^3 i$ (Table~2), 
are quite large, and given the relative scarcity of data on 
the masses of O-type stars \citep{bur97}, it is important 
to compare our results with predictions based on evolutionary 
tracks.  Since HD~101131 is a member of the cluster IC~2944, 
we can use the cluster distance modulus to help place the 
individual component stars in the H-R diagram. 
We estimated the stellar temperatures from the spectral 
classifications using the calibrated sequence given by 
\citet{how89}.   These temperatures and other stellar parameters 
are listed in Table~3.   
We used flux models from \citet{kur94} to transform our observed
flux ratio based upon the relative line depths in the blue part 
of the spectrum into a $V$-band flux ratio \citep{pen97}, 
$f_2(V)/f_1(V) = 0.55\pm 0.08$, so that we 
could determine the $V$ magnitude of both components. 
We adopted the composite magnitude, $V = 7.16$, and cluster distance modulus, 
$m-M = 12.5 \pm 0.2$, from \citet{tha65}, which agree well with more recent 
estimates \citep{tov98}.  We used the bolometric corrections
given by \citet{how89} to arrive at estimates of the individual 
luminosities, and these are listed in Table~3.   Table~3 also includes 
the resulting stellar radii estimates.   We calculated 
the sizes of the Roche volume radius \citep{egg83} 
of both stars for the time of smallest separation (Table~3), and both are well 
within their critical Roche surfaces then, even after accounting 
for their rapid rotation \citep{lim63}.  Thus, given these 
stellar dimensions and the youthful appearance of the IC~2944 cluster, 
it is safe to assume that the interacting stage in this system 
lies far in the future and that we can compare the
stars' evolutionary state with the predictions for single stars
with the same physical parameters.

\placetable{tab3}      

The primary and secondary components are placed in the H-R diagram in 
Figure~3 together with the evolutionary tracks for single stars
from the models of \citet{sch92}.   Both stars appear close to 
the zero-age main sequence, and an interpolation in these tracks 
gives ages of $1.5\pm 0.4$ and $2.7\pm 0.8$ million years for 
the primary and secondary, respectively.  The predicted current 
masses from these evolutionary tracks are listed in Table~3.   
Note that both stars are relatively rapid rotators (with 
angular velocities greater than 2 and 4 times synchronous 
for the primary and secondary, respectively).
Rapidly rotating stars may have luminosities larger 
than their nonrotating counterparts \citep{heg00,mey00}, so 
the predicted masses given in Table~2 may be slight 
overestimates.  

\placefigure{fig3}     

Lower limits for the masses can be taken directly from 
the $M \sin^3 i$ values, but we can improve these somewhat
from light-curve considerations.   The target was observed
by the {\it Hipparcos} satellite \citep{per97}, and we 
examined the {\it Hipparcos} epoch photometry data for any evidence of
eclipses based on our orbital ephemeris.   There is no 
sign of any eclipse deeper than 0.01 mag, so we 
can use the lack of eclipses together with our estimates 
of the binary separation at superior conjunction of the 
primary (the closer of the two conjunctions) and the stellar
radii to find an upper limit on the orbital inclination, 
$i < (72\fdg0 \pm 1\fdg6)$.   The resulting lower limits 
for the stellar masses are listed in the final row of Table~3. 
We find that the predicted masses are somewhat larger than 
these lower limits, and, in fact, if the evolutionary masses 
are correct, then the system inclination is actually $i = 56^\circ
\pm 2^\circ$.   A second consistency check is obtained 
from the implied mass ratio from the evolutionary models, 
$q= M_2/M_1 = 0.63\pm 0.05$, that agrees within errors with 
the observed mass ratio, $q= 0.56\pm 0.02$.  Thus, 
we find that the constraints on the system masses provided 
by the orbital elements and distance modulus are fully   
consistent with the predictions from evolutionary tracks,
in agreement with the conclusions of \citet{bur97} for 
other unevolved close binary systems.   The predicted maximum
angular separation is 0.15~mas, which may be within the reach of 
future optical interferometers. 


\acknowledgments

We thank the staffs of CASLEO and MSO for their assistance 
in making these observations.
We also thank Alex Fullerton for sharing his period
search code with us.  We are grateful to
Dr.\ T.\ Lloyd Evans (SAAO), who kindly arranged the plate loan, 
and to Mr.\ J.\ Havelka (Astronomical Institute, Czech Academy of Sciences) 
who digitized these plates.  Institutional support for L.R.P.\
has been provided from the College of Charleston School
of Sciences and Mathematics.  Additional support for L.R.P.\ was
provided by the South Carolina NASA Space Grant Program and
NSF grant AST-9528506.
Institutional support for D.R.G.\ has been provided from the GSU College
of Arts and Sciences and from the Research Program Enhancement
fund of the Board of Regents of the University System of Georgia,
administered through the GSU Office of the Vice President
for Research.  We gratefully acknowledge all this support.




\clearpage

\begin{deluxetable}{lccccrcccr}
\tabletypesize{\scriptsize}
\tablewidth{0pt}
\tablenum{1}
\tablecaption{Radial Velocity Measurements \label{tab1}}
\tablehead{
\colhead{HJD}             &
\colhead{Orbital}         &
\colhead{$V_1$}           &
\colhead{$\sigma_1$}      &
\colhead{$(O-C)_1$}       &
\colhead{}                &
\colhead{$V_2$}           &
\colhead{$\sigma_2$}      &
\colhead{$(O-C)_2$}       &
\colhead{}                \\
\colhead{(-2,400,000)} &
\colhead{Phase}           &
\colhead{(km s$^{-1}$)}   &
\colhead{(km s$^{-1}$)}   &
\colhead{(km s$^{-1}$)}   &
\colhead{$n_1$}           &
\colhead{(km s$^{-1}$)}   &
\colhead{(km s$^{-1}$)}   &
\colhead{(km s$^{-1}$)}   &
\colhead{$n_2$}           }
\scriptsize
\startdata
 34060.492\tablenotemark{a} \dotfill &  0.570 & 
\phs         $ 106.0$ &\nodata &\phs     $  29.9$ &  6 &
             $-155.0$ &\nodata &         $ -20.4$ &  3 \\
 34086.394 \dotfill &  0.255 & 
\phn         $ -60.0$ &\nodata &\phs     $  27.6$ & \nodata &
\nodata               &\nodata &\nodata           & \nodata  \\
 34094.377\tablenotemark{a} \dotfill &  0.083 & 
             $-121.0$ &\nodata &\phn\phs $   5.1$ &  6 &
\phs         $ 206.0$ &\nodata &         $ -23.6$ &  2 \\
 34107.349\tablenotemark{a} \dotfill &  0.427 & 
\phn\phs     $  27.0$ &\nodata &\phs     $  16.0$ &  5 &
\nodata               &\nodata &\nodata           & \nodata  \\
 34114.380\tablenotemark{a} \dotfill &  0.156 & 
             $-134.0$ &\nodata &\phn     $  -6.3$ &  3 &
\nodata               &\nodata &\nodata           & \nodata  \\
 34144.298\tablenotemark{a} \dotfill &  0.258 & 
             $-117.0$ &\nodata &         $ -30.8$ &  7 &
\phs         $ 153.0$ &\nodata &\phn     $  -4.8$ &  2 \\
 34179.210\tablenotemark{a} \dotfill &  0.877 & 
\phn         $ -25.0$ &\nodata &         $ -62.9$ &  6 &
\nodata               &\nodata &\nodata           & \nodata  \\
 34193.192\tablenotemark{a} \dotfill &  0.326 & 
\phn         $ -12.0$ &\nodata &\phs     $  35.0$ &  7 &
\nodata               &\nodata &\nodata           & \nodata  \\
 34768.580 \dotfill &  0.973 & 
\phn\phs     $  16.0$ &\nodata &\phs     $  67.9$ & \nodata &
\nodata               &\nodata &\nodata           & \nodata  \\
 35594.308 \dotfill &  0.571 & 
\phs         $ 128.0$ &\nodata &\phs     $  51.6$ & \nodata &
\nodata               &\nodata &\nodata           & \nodata  \\
 36595.565 \dotfill &  0.365 & 
\phn         $ -34.0$ &\nodata &\phn     $  -9.6$ & \nodata &
\nodata               &\nodata &\nodata           & \nodata  \\
 36598.533 \dotfill &  0.673 & 
\phn\phs     $  56.0$ &\nodata &         $ -44.6$ & \nodata &
\nodata               &\nodata &\nodata           & \nodata  \\
 36599.472 \dotfill &  0.770 & 
\phn\phs     $  37.0$ &\nodata &         $ -57.4$ & \nodata &
\nodata               &\nodata &\nodata           & \nodata  \\
 36602.499 \dotfill &  0.084 & 
             $-104.0$ &\nodata &\phs     $  22.3$ & \nodata &
\nodata               &\nodata &\nodata           & \nodata  \\
 36604.493 \dotfill &  0.290 & 
\phn         $ -64.0$ &\nodata &\phn\phs $   4.0$ & \nodata &
\nodata               &\nodata &\nodata           & \nodata  \\
 36604.564 \dotfill &  0.298 & 
\phn         $ -14.0$ &\nodata &\phs     $  49.7$ & \nodata &
\nodata               &\nodata &\nodata           & \nodata  \\
 36605.476 \dotfill &  0.392 & 
\phn\phs     $  12.0$ &\nodata &\phs     $  20.6$ & \nodata &
\nodata               &\nodata &\nodata           & \nodata  \\
 36613.497 \dotfill &  0.224 & 
\phn         $ -65.0$ &\nodata &\phs     $  38.5$ & \nodata &
\nodata               &\nodata &\nodata           & \nodata  \\
 36624.547 \dotfill &  0.369 & 
\phn\phn     $  -4.0$ &\nodata &\phs     $  17.8$ & \nodata &
\nodata               &\nodata &\nodata           & \nodata  \\
 38844.402 \dotfill &  0.487 & 
\phn\phn     $  -9.0$ &\nodata &         $ -50.6$ & \nodata &
\nodata               &\nodata &\nodata           & \nodata  \\
 38858.343 \dotfill &  0.933 & 
\phn\phs     $  17.0$ &\nodata &\phs     $  29.3$ & \nodata &
\nodata               &\nodata &\nodata           & \nodata  \\
 39180.472 \dotfill &  0.326 & 
\phn\phn\phs $   2.0$ &\nodata &\phs     $  49.4$ & \nodata &
\nodata               &\nodata &\nodata           & \nodata  \\
 39180.578 \dotfill &  0.337 & 
\phn\phs     $  26.0$ &\nodata &\phs     $  66.9$ & \nodata &
\nodata               &\nodata &\nodata           & \nodata  \\
 39186.428 \dotfill &  0.943 & 
\phn\phs     $  25.0$ &\nodata &\phs     $  47.6$ & \nodata &
\nodata               &\nodata &\nodata           & \nodata  \\
 39186.521 \dotfill &  0.953 & 
\phn         $ -11.0$ &\nodata &\phs     $  21.0$ & \nodata &
\nodata               &\nodata &\nodata           & \nodata  \\
 39207.351 \dotfill &  0.112 & 
             $-121.0$ &\nodata &\phs     $  10.2$ & \nodata &
\nodata               &\nodata &\nodata           & \nodata  \\
 39207.465 \dotfill &  0.124 & 
             $-114.0$ &\nodata &\phs     $  17.5$ & \nodata &
\nodata               &\nodata &\nodata           & \nodata  \\
 39252.231 \dotfill &  0.764 & 
\phn\phs     $  56.0$ &\nodata &         $ -39.8$ & \nodata &
\nodata               &\nodata &\nodata           & \nodata  \\
 39269.208 \dotfill &  0.524 & 
\phn\phn     $  -8.0$ &\nodata &         $ -66.3$ & \nodata &
\nodata               &\nodata &\nodata           & \nodata  \\
 39518.485 \dotfill &  0.365 & 
\phn         $ -24.0$ &\nodata &\phn\phs $   0.2$ & \nodata &
\nodata               &\nodata &\nodata           & \nodata  \\
 39518.578 \dotfill &  0.375 & 
\phn         $ -14.0$ &\nodata &\phn\phs $   4.6$ & \nodata &
\nodata               &\nodata &\nodata           & \nodata  \\
 39556.424 \dotfill &  0.298 & 
\phn         $ -58.0$ &\nodata &\phn\phs $   5.5$ & \nodata &
\nodata               &\nodata &\nodata           & \nodata  \\
 39610.306 \dotfill &  0.884 & 
\phn\phn     $  -5.0$ &\nodata &         $ -37.3$ & \nodata &
\nodata               &\nodata &\nodata           & \nodata  \\
 39625.397 \dotfill &  0.448 & 
\phn         $ -10.0$ &\nodata &         $ -32.0$ & \nodata &
\nodata               &\nodata &\nodata           & \nodata  \\
 39655.205 \dotfill &  0.538 & 
\phn\phs     $  38.0$ &\nodata &         $ -26.0$ & \nodata &
\nodata               &\nodata &\nodata           & \nodata  \\
 39655.304 \dotfill &  0.548 & 
\phn\phs     $  48.0$ &\nodata &         $ -20.1$ & \nodata &
\nodata               &\nodata &\nodata           & \nodata  \\
 40696.285 \dotfill &  0.460 & 
\phn\phs     $  16.0$ &\nodata &         $ -12.2$ & \nodata &
\nodata               &\nodata &\nodata           & \nodata  \\
 40696.492 \dotfill &  0.482 & 
\phn\phn\phs $   6.0$ &\nodata &         $ -32.8$ & \nodata &
\nodata               &\nodata &\nodata           & \nodata  \\
 40697.295 \dotfill &  0.565 & 
\phn\phs     $  21.0$ &\nodata &         $ -53.3$ & \nodata &
\nodata               &\nodata &\nodata           & \nodata  \\
 40697.469 \dotfill &  0.583 & 
\phn\phs     $  51.0$ &\nodata &         $ -29.5$ & \nodata &
\nodata               &\nodata &\nodata           & \nodata  \\
 40698.295 \dotfill &  0.669 & 
\phs         $ 101.0$ &\nodata &\phn\phs $   0.9$ & \nodata &
\nodata               &\nodata &\nodata           & \nodata  \\
 40698.469 \dotfill &  0.687 & 
\phs         $ 101.0$ &\nodata &\phn     $  -0.8$ & \nodata &
\nodata               &\nodata &\nodata           & \nodata  \\
 42119.780\tablenotemark{b} \dotfill &  0.025 & 
\phn         $ -71.3$ & 3.9    &\phs     $  25.1$ & 22 &
\nodata               &\nodata &\nodata           & \nodata  \\
 44258.512\tablenotemark{c} \dotfill &  0.734 & 
\phs         $ 121.4$ &\nodata &\phs     $  20.4$ &  1 &
             $-136.1$ &\nodata &\phs     $  43.3$ &  1 \\
 48676.621 \dotfill &  0.731 & 
\phs         $ 157.0$ &\nodata &\phs     $  55.7$ &  1 &
             $-124.0$ &\nodata &\phs     $  55.9$ &  1 \\
 48678.679 \dotfill &  0.944 & 
\phn\phs     $  38.0$ &\nodata &\phs     $  61.4$ &  1 &
\nodata               &\nodata &\nodata           & \nodata  \\
 48679.666 \dotfill &  0.046 & 
             $-100.0$ &\nodata &\phs     $  10.3$ &  1 &
\phs         $ 120.0$ &\nodata &         $ -81.2$ &  1 \\
 48763.596 \dotfill &  0.747 & 
\phn\phs     $  61.0$ &\nodata &         $ -38.3$ &  4 &
             $-122.0$ &\nodata &\phs     $  54.3$ &  4 \\
 49028.717 \dotfill &  0.230 & 
             $-118.0$ &\nodata &         $ -17.5$ &  4 &
\phs         $ 196.0$ &\nodata &\phs     $  12.5$ &  4 \\
 49151.548 \dotfill &  0.963 & 
\phn\phn     $  -2.0$ &\nodata &\phs     $  40.3$ &  2 &
\nodata               &\nodata &\nodata           & \nodata  \\
 49451.763 \dotfill &  0.085 & 
\phn         $ -81.0$ &\nodata &\phs     $  45.6$ &  1 &
\phs         $ 206.0$ &\nodata &         $ -24.5$ &  1 \\
 49454.570 \dotfill &  0.376 & 
\phn         $ -56.0$ &\nodata &         $ -37.8$ &  1 &
\nodata               &\nodata &\nodata           & \nodata  \\
 50526.791 \dotfill &  0.526 & 
\phn\phs     $  68.4$ & 3.6    &\phn\phs $   9.5$ & 15 &
             $-148.3$ &\phn 5.2&         $ -44.7$ & 15 \\
 50527.749 \dotfill &  0.625 & 
\phn\phs     $  92.9$ & 0.9    &\phn\phs $   0.6$ & 11 &
             $-150.6$ &\phn 4.0&\phs     $  13.3$ & 15 \\
 50528.796 \dotfill &  0.734 & 
\phs         $ 118.9$ & 1.4    &\phs     $  17.9$ & 14 &
             $-188.7$ &\phn 4.5&\phn     $  -9.3$ & 13 \\
 50529.763 \dotfill &  0.834 & 
\phn\phs     $  78.0$ & 3.5    &\phs     $  10.4$ & 15 &
             $-110.6$ &\phn 9.4&\phn\phs $   8.7$ & 12 \\
 50530.747 \dotfill &  0.936 & 
\phn         $ -40.5$ & 8.2    &         $ -25.0$ & 14 &
\phn\phs     $  23.8$ & 22.3   &\phn     $  -6.7$ & 15 \\
 50531.724 \dotfill &  0.037 & 
             $-109.2$ & 5.3    &\phn     $  -4.3$ & 15 &
\phs         $ 158.2$ & 12.0   &         $ -33.1$ & 12 \\
 50532.753 \dotfill &  0.144 & 
             $-132.7$ & 4.5    &\phn     $  -2.8$ & 13 &
\phs         $ 215.0$ &\phn 7.0&         $ -21.5$ & 13 \\
 50533.724 \dotfill &  0.245 & 
\phn         $ -77.2$ & 3.1    &\phs     $  16.0$ & 14 &
\phs         $ 149.0$ &\phn 3.6&         $ -21.4$ & 12 \\
 50534.758 \dotfill &  0.352 & 
\phn         $ -46.5$ & 7.4    &         $ -14.4$ & 15 &
\phn\phs     $  64.9$ & 11.0   &\phn\phs $   4.5$ & 11 \\
 50535.737 \dotfill &  0.453 & 
\phn\phn\phs $   6.0$ & 4.9    &         $ -18.5$ & 15 &
\phn\phs     $  14.7$ & 10.7   &\phs     $  56.4$ & 15 \\
 50909.952 \dotfill &  0.246 & 
\phn         $ -98.1$ & 2.7    &\phn     $  -5.5$ &  7 &
\phs         $ 193.8$ &\phn 7.0&\phs     $  24.4$ &  7 \\
 50910.190 \dotfill &  0.270 & 
\phn         $ -88.7$ & 1.8    &\phn     $  -9.4$ &  7 &
\phs         $ 168.5$ &\phn 2.1&\phs     $  23.1$ &  6 \\
 50910.935 \dotfill &  0.348 & 
\phn         $ -26.1$ & 6.3    &\phn\phs $   8.4$ &  7 &
\phn\phs     $  40.7$ & 22.8   &         $ -24.0$ &  7 \\
 50911.038 \dotfill &  0.358 & 
\phn         $ -16.5$ & 6.8    &\phs     $  11.8$ &  7 &
\phn\phs     $  25.3$ & 19.9   &         $ -28.1$ &  7 \\
 50914.179 \dotfill &  0.684 & 
\phs         $ 112.6$ & 2.8    &\phs     $  11.0$ &  7 &
             $-188.1$ &\phn 5.7&\phn     $  -7.6$ &  7 \\
 50914.952 \dotfill &  0.764 & 
\phn\phs     $  97.7$ & 1.0    &\phn\phs $   1.8$ &  6 &
             $-165.5$ &\phn 9.0&\phn\phs $   4.7$ &  7 \\
 50916.148 \dotfill &  0.888 & 
\phn\phn     $  -0.4$ & 9.5    &         $ -29.2$ &  7 &
\nodata               &\nodata &\nodata           & \nodata  \\
 50916.913 \dotfill &  0.967 & 
\phn         $ -28.8$ & 4.6    &\phs     $  17.5$ &  7 &
\phn\phs     $  98.2$ &\phn 5.6&\phs     $  12.3$ &  6 \\
 50917.060 \dotfill &  0.983 & 
\phn         $ -48.3$ & 1.3    &\phs     $  12.5$ &  6 &
\phs         $ 102.2$ &\phn 3.7&\phn     $  -9.9$ &  6 \\
 50917.204 \dotfill &  0.997 & 
\phn         $ -68.8$ & 1.2    &\phn\phs $   5.5$ &  6 &
\phs         $ 125.4$ &\phn 1.8&         $ -11.1$ &  6 \\
 50917.936 \dotfill &  0.073 & 
             $-136.9$ & 2.2    &         $ -13.9$ &  7 &
\phs         $ 252.4$ &\phn 9.7&\phs     $  28.4$ &  7 \\
 50918.058 \dotfill &  0.086 & 
             $-140.1$ & 1.8    &         $ -13.2$ &  7 &
\phs         $ 253.4$ &\phn 8.5&\phs     $  22.2$ &  7 \\
 50918.217 \dotfill &  0.102 & 
             $-136.4$ & 2.5    &\phn     $  -6.2$ &  7 &
\phs         $ 255.5$ &\phn 9.4&\phs     $  18.4$ &  7 \\
\enddata
\tablenotetext{a}{Feast et al.\ (1955)}
\tablenotetext{b}{Conti et al.\ (1977)}
\tablenotetext{c}{Stickland \& Lloyd (2001)}
\end{deluxetable}

\newpage

\begin{deluxetable}{lr}
\tablewidth{0pc}
\tablenum{2}
\tablecaption{Orbital Elements  \label{tab2}}
\tablehead{
\colhead{Element} & 
\colhead{Value} }
\startdata
$P$~(days)                  \dotfill & 9.64659 (12) \\
$T$ (HJD-2,400,000)         \dotfill & 48650.28 (29) \\
$e$                         \dotfill & 0.156 (29) \\
$\omega$ ($^\circ$)         \dotfill & 122 (12) \\
$K_1$ (km s$^{-1}$)         \dotfill & 117 (4) \\
$K_2$ (km s$^{-1}$)         \dotfill & 211 (7) \\
$V_{0~1}$ (km s$^{-1}$)     \dotfill & $-4.9$ (25) \\
$V_{0~2}$ (km s$^{-1}$)     \dotfill & 11 (5) \\
$M_1$ sin$^{3}i$ ($M_\odot$)\dotfill & 21.8 (21) \\
$M_2$ sin$^{3}i$ ($M_\odot$)\dotfill & 12.1 (13) \\
$a_1$ sin $i$ ($R_\odot$)   \dotfill & 22.0 (7) \\
$a_2$ sin $i$ ($R_\odot$)   \dotfill & 39.7 (12) \\
r.m.s.$_1$ (km s$^{-1}$)    \dotfill & 20.6 \\
r.m.s.$_2$ (km s$^{-1}$)    \dotfill & 25.4 \\
\enddata
\tablecomments{Numbers in parentheses give the error in the last digit quoted.}
\end{deluxetable}

\newpage

\begin{deluxetable}{lcc}
\tablewidth{0pc}
\tablenum{3}
\tablecaption{Stellar Properties \label{tab3}}
\tablehead{
\colhead{Property} & 
\colhead{Primary} &
\colhead{Secondary} }
\startdata
Spectral Classification              \dotfill &  O6.5~V((f))   &  O8.5~V        \\
Relative flux $F/F_1$(4300\AA )      \dotfill &  1.0           & $0.54\pm 0.08$ \\
$V\sin i$ (km s$^{-1}$)              \dotfill & $102\pm 10$    &  $164\pm 12$   \\
$T_{\rm eff}$ (kK)                   \dotfill & $40.5\pm 1.5$  & $35.0\pm 1.5$  \\
$\log L/L_\odot$                     \dotfill & $5.34\pm 0.08$ & $4.91\pm 0.09$ \\
$R/R_\odot$                          \dotfill &  $9.5\pm 0.9$  &  $7.7\pm 0.8$  \\
$(R_{\rm Roche} \sin i) /R_\odot$ (periastron) \dotfill & $22.4\pm 1.0$  & $17.1\pm 0.7$  \\
$M/M_\odot$ (predicted)              \dotfill & $36.2\pm 2.0$  & $22.9\pm 1.1$  \\
$M/M_\odot$ (observed limit)         \dotfill & $>25.3\pm 2.5$ & $>14.1\pm 1.5$ \\
\enddata
\end{deluxetable}



\clearpage

\begin{figure}
\caption{
The radial velocity measurements 
({\it filled circles}: CASLEO and MSO data for the primary; 
{\it plus signs}: low weight data for the primary;
{\it open circles}: CASLEO and MSO data for the secondary;  
{\it X signs}: low weight data for the secondary) 
and orbital solution ({\it solid lines}) plotted against
orbital phase.  Phase zero corresponds to the time of periastron.
}
\label{fig1}
\end{figure}

\begin{figure}
\caption{
Comparisons of the reconstructed primary spectra 
(MSO offset to a continuum of 2.8; CASLEO offset to a continuum of 2.5) 
and reconstructed secondary spectra 
(MSO offset to a continuum of 1.3; CASLEO at the normal continuum of 1.0)
with spectra of the same classifications 
from \citet{wal90}.  All the spectra were Gaussian smoothed 
to a nominal resolution of 1.2 \AA ~FWHM for comparable line broadening. }
\label{fig2}
\end{figure}

\begin{figure}
\caption{
H-R diagram for the components of HD 101131 ({\it filled circle}: primary;
{\it open circle}: secondary).  {\it Solid lines}: single star evolutionary 
tracks of \citet{sch92} labelled by the initial stellar mass ($M_\odot$). 
{\it Diagonal error lines}: range in effective temperature and bolometric 
correction associated with the spectral classification. 
{\it Vertical error bars}: net error in luminosity for the adopted temperature.
Our derived lower limits on the masses (25 and $14 M_\odot$ for the primary 
and secondary, respectively) are consistent with the predicted evolutionary masses.
}
\label{fig3}
\end{figure}



\clearpage

\setcounter{figure}{0}

\begin{figure}
\plotone{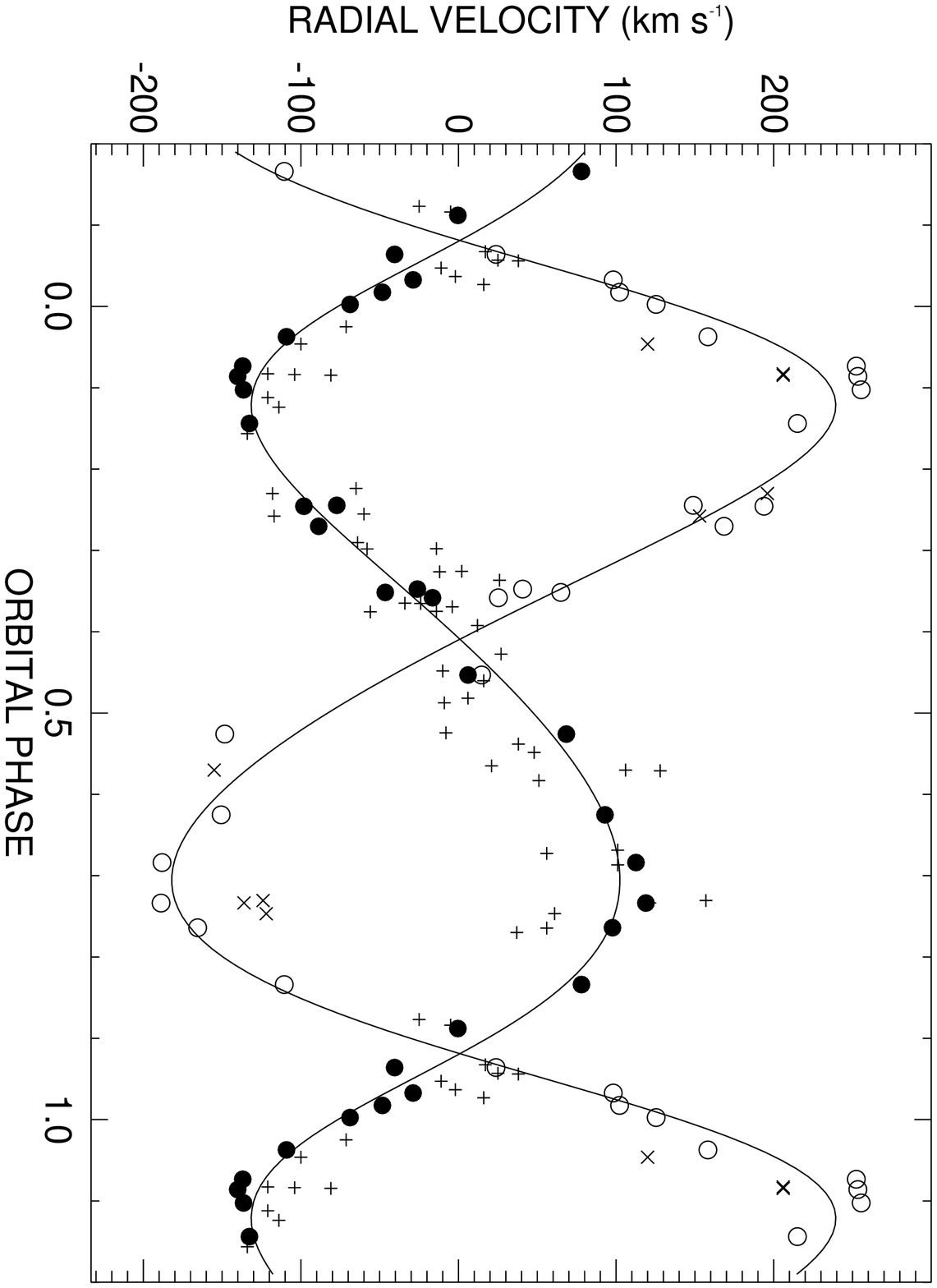}
\caption{}
\end{figure}

\begin{figure}
\plotone{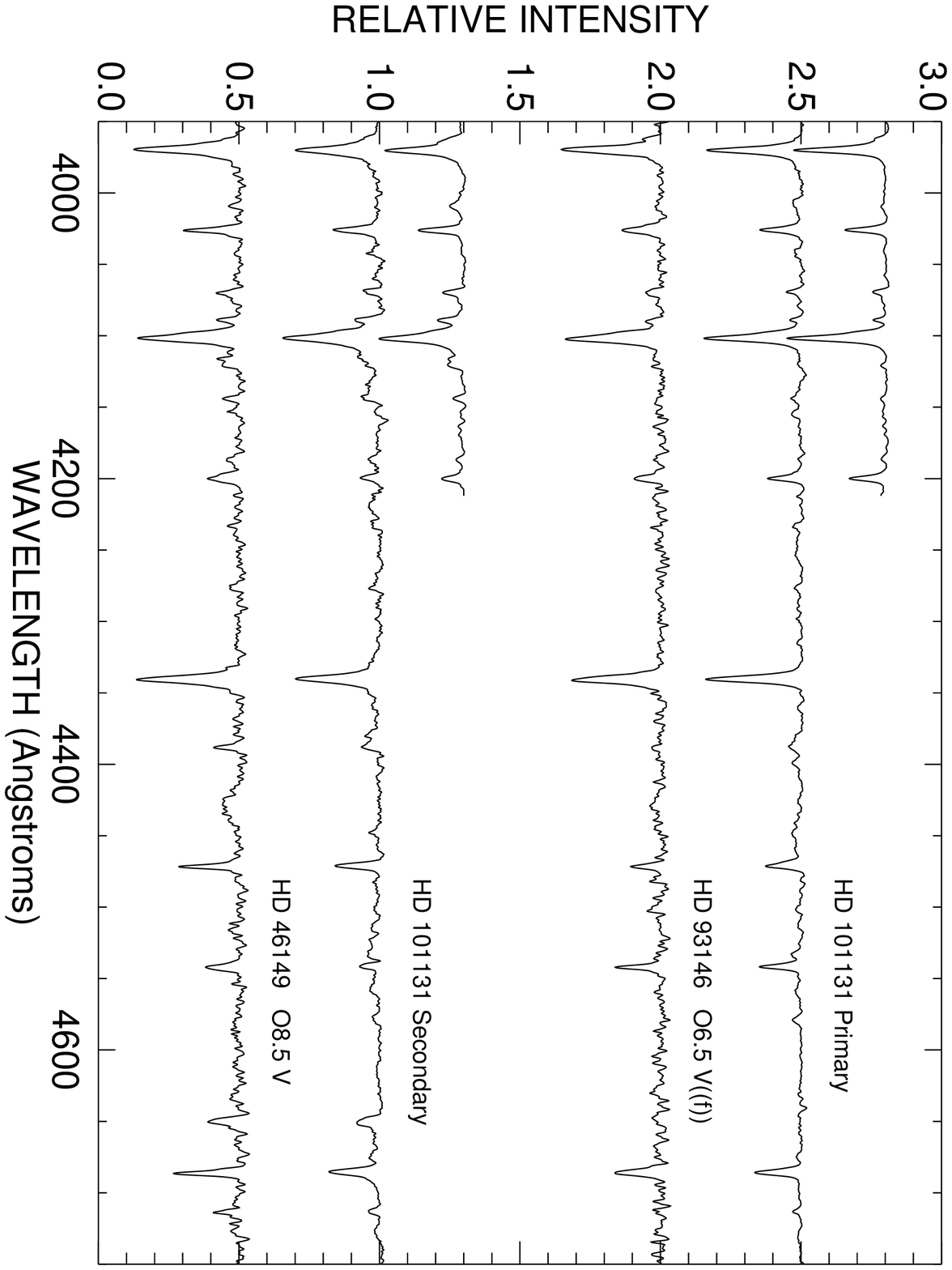}
\caption{}
\end{figure}

\begin{figure}
\plotone{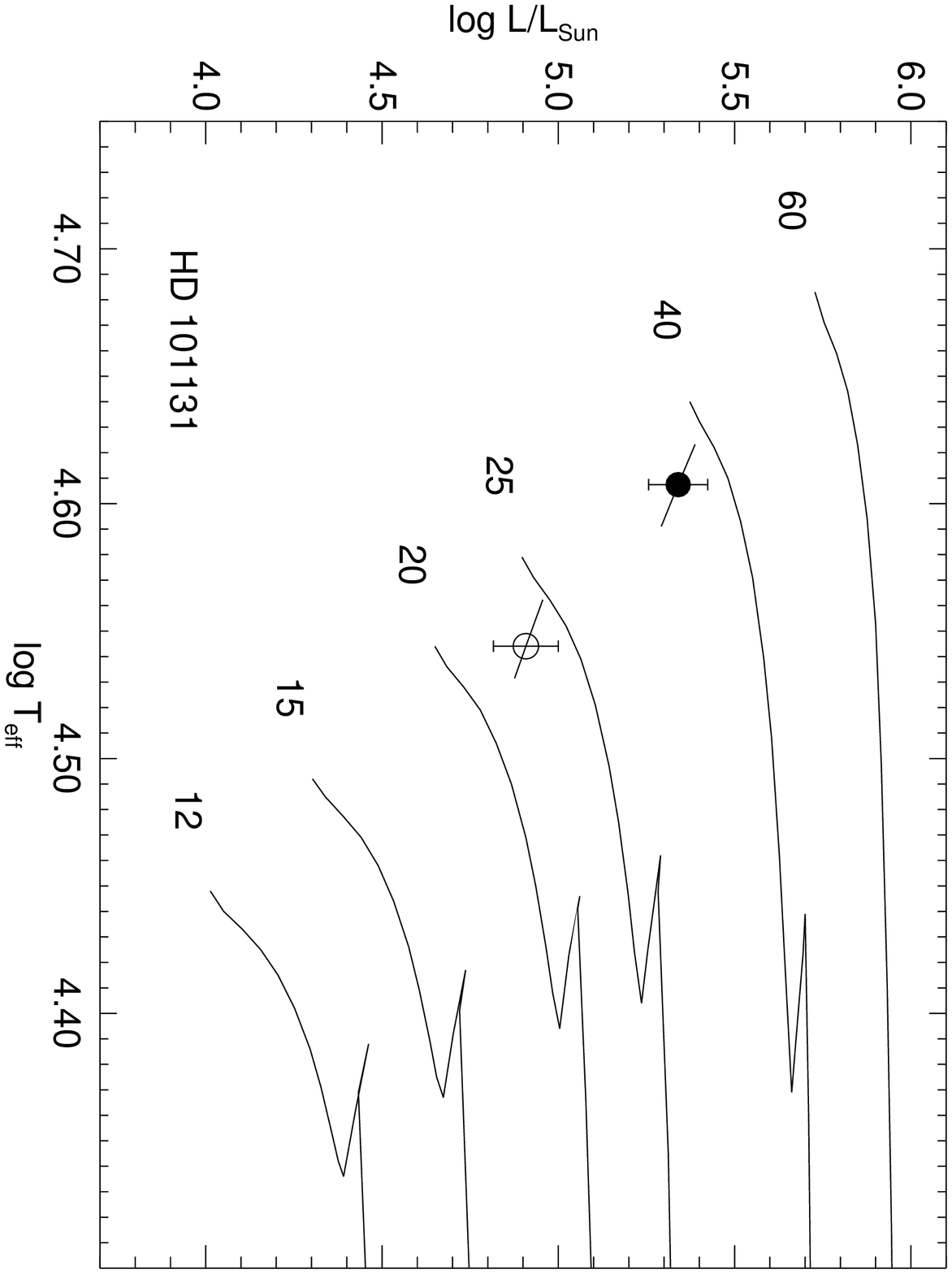}
\caption{}
\end{figure}


\end{document}